# Multisensory cues facilitate coordination of stepping movements with a virtual reality avatar

Omar Khan[1,¶], Imran Ahmed[2,&], Joshua Cottingham[3,&], Musa Rahhal[4,&], Theodoros N. Arvanitis[1,¶], Mark T. Elliott[1,¶,*]

[1] Institute of Digital Healthcare, WMG, University of Warwick, Coventry, UK

[2] Warwick Medical School, University of Warwick, Coventry, UK

[3] Department of Computer Science, University of Warwick, Coventry, UK

[4] School of Engineering, University of Warwick, Coventry, UK

\* Corresponding author

E-mail: m.t.elliott@warwick.ac.uk (MTE)

[¶]These authors contributed equally to this work.

[&]These authors also contributed equally to this work.

The effectiveness of simple sensory cues for retraining gait have been demonstrated, yet the feasibility of humanoid avatars for entrainment have yet to be investigated. Here, we describe the development of a novel method of visually cued training, in the form of a virtual partner, and investigate its ability to provide movement guidance in the form of stepping.

Real stepping movements were mapped onto an avatar using motion capture data. The trajectory of one of the avatar step cycles was then accelerated or decelerated by 15% to create a perturbation. Healthy participants were motion captured while instructed to step in time to the avatar's movements, as viewed through a virtual reality headset. Step onset times were used to measure the timing errors (asynchronies) between them. Participants completed either a visual-only condition, or auditory-visual with footstep sounds included.

Participants' asynchronies exhibited slow drift in the Visual-Only condition, but became stable in the Auditory-Visual condition. Moreover, we observed a clear corrective response to the phase perturbation in both auditory-visual conditions.

We conclude that an avatar's movements can be used to influence a person's own gait, but should include relevant auditory cues congruent with the movement to ensure a suitable accuracy is achieved.

# Introduction

Mirroring of movements has been described as an important aspect of social interactions[1], as well as providing a top-down process that allows humans to anticipate actions and their goals when interacting with others[2]. This process involves the mapping of the visual information about an agent's movement on to one's own interpreted motor representation[3]. The opportunity to exploit action mirroring, for movement training, is clear and has been investigated in the context of observing familiar and unfamiliar dance moves[4] and, in early experimental work, as visual cues for gait retraining[5].

An important area of movement training is in the context of physiotherapy and rehabilitation. Following lower limb or neurological injury, gait retraining is often required as an intense part of a physiotherapy or rehabilitation process. In the rehabilitation of gait function, patients are set goal-oriented functional tasks, such as walking, turning and stepping over objects[6], depending on the nature of their impairment. Commonly, these tasks fall into three main categories: neurodevelopmental techniques, strength training and task-specific training, including treadmill and intensive mobility training[7]. In the hospital setting, these are normally performed in the presence of physio- or occupational-therapists, who provide feedback to the patient and update the training regimen as required[8]. Outside the clinic, patients are usually left with limited guidance on their exercise routine. Subsequently, patients can quickly lose motivation or complete exercises incorrectly without the presence of physiotherapist supervision[9]. Here, we investigate the feasibility of an avatar to provide movement guidance to participants, by investigating how accurately participants can coordinate their movements with the avatar.

Simple auditory and visual cues have been used in gait retraining during neuro-rehabilitation programmes[10]. For example, using an auditory metronome to provide a regular beat has been found to be an effective method for retraining the timing and coordination of gait following Stroke[11] or Parkinson's disease[12]. Similarly, the use of regularly timed visual stepping stones on a treadmill can be used to influence gait coordination[13]. However, the influence exerted by auditory cues versus visual cues is disputed. In some contexts, (e.g., side by side walking[14]) the visual component was found to have less of an influence on step synchronisation than the auditory component. In contrast, the use of visual stepping stones on treadmill revealed that corrections to visual cues were faster than auditory cues, thus indicating gait may be more strongly influenced by visual cues in certain contexts[13]. The visual stimuli used in the majority of the published literature to date has been limited to simple representations, such as flashing lights or bouncing spheres. The abstract nature of the cues used may limit the scope of their effect. This is highlighted by the finding that maintaining a defined distance to an avatar was more accurate compared to the same task using simple spheres[15]. In the temporal domain, synchronisation of movements to spatio-temporal visual stimuli is more accurate than to just temporal stimuli[16],[17],[18]. The presence of spatio-temporal information along with a possibility to invoke the mirror system, through representation of familiar movements, suggests an avatar could provide effective cues for gait retraining, as indicated by the effectiveness of point-light figures in inducing beat entrainment over auditory beats[19].

In this study, we investigate how accurately participants are able to coordinate stepping movements with an avatar in a virtual reality environment through the use of a sensorimotor synchronisation paradigm. A fully immersive environment was created using a VR headset, such that participants could view the avatar in three-dimensions, including the ability to move around relative to the avatar's position. Following studies that have shown that multisensory cues can improve the accuracy of sensorimotor synchronisation[20,21], the avatar was presented in either a visual-only format (no contextual audio cues relating to foot strikes) or visually with contextual auditory footsteps. Audio cueing experiments have already provided clear evidence of the correction process, in both finger tapping and stepping, and have demonstrated evidence that multisensory integration takes place with visual cues. Hence, our focus in this study was the response to humanoid visual stimulus comprising temporal and spatial cues.

To test whether participants were truly synchronising with the avatar (rather than just generating 'similar' responses), we used a phase perturbation paradigm, by lengthening or shortening one of gait cycles during a trial by an amount that was unnoticeable by participants. Subsequent changes in asynchrony between the heel-strikes of participants relative to the avatar were analysed to determine if participants corrected their timing to get back in synchrony following the perturbation. Overall, we hypothesised that 1) Participants would be able to match the tempo of their steps with the Avatar's, at both fast and slow stepping tempos; 2) Participants would be able to synchronise their steps with the Avatar's and subsequently correct their steps following a perturbation, but would exhibit greater accuracy (lower asynchrony variance and faster correction) when the avatar's movement was augmented with auditory footsteps.

# Results

Participants wore a VR headset, in which they saw an avatar stood approximately 3m in front of them. The avatar was facing in the same direction as the participant, such that the participant was viewing the back of the avatar. Participants were fitted with a full body marker set (Plug-in Gait, Vicon[22]), such that their own movements could be accurately tracked. They were instructed to step in time with the avatar as accurately as they could. Each trial included a perturbation to a single gait cycle of the avatar, equivalent to 15% of the inter-step interval. Participants completed either Experiment 1 (Visual-only) or Experiment 2 (Audio-visual), which involved 4 blocks of 5 trials (20 trials in total). For each between-subjects condition, there were within-subjects conditions of Tempo: Slow (800ms inter-step interval), Fast (400ms inter-step interval) and Perturbation Direction: Positive (longer gait cycle interval), Negative (shorter gait cycle interval).

## Matching of inter-step interval

To evaluate whether participants were able to match their stepping tempo to the avatar, the mean inter-step interval (ISI) for each participant per trial type was calculated (A mixed measures ANOVA was used to compare mean and standard deviation of participant ISIs across conditions. In addition, ). To calculate this value, steps onsets were calculated as the point the heel marker exceeded a defined vertical height (see Methods; Fig 1a). The intervals were calculated as the time between the current and preceding step onset. Intervals were then averaged between the third step interval and the interval preceding the perturbed step.

**Fig 1. Example step onsets and resulting temporal performance measures a).** Example of step onsets (red circles) extracted from the vertical heel marker trajectory (blue line) during the stepping task. Onsets were extracted from the avatar and participants in the same way. **b**). Schematic showing temporal performance measures based on the step onset times of the avatar and participant. Asynchrony ($A$) was measured as the time difference between the participant's and corresponding avatar step onsets. Inter-step intervals for the avatar ($I^{Av}$) and participant ($I^P$) were calculated as the time between a participant's/avatar's own step onsets. Step $T$, referred to the avatar step onset that was perturbed (shown here as a shortened interval by shifting the onset earlier).

A mixed measures ANOVA was used to compare mean and standard deviation of participant ISIs across conditions. In addition, each participant's mean ISI for each condition was compared to the avatar's corresponding mean ISI using a paired t-test (with adjusted p-value for multiple comparisons). Due to the Avatar movements being based on real movement data (see Methods) and slight variations in frame rate, the mean Avatar ISI also varied slightly across participants and trials.

On the whole, participants were clearly able to differentiate between the two avatar ISI conditions (400ms and 800ms) and adjust their own step timing accordingly ($F_{1,18}=719.169, p<.001$; **Error! Reference source not found.**). This indicates that the avatar was successfully able to influence participants to step at a pace other than a self-selected ISI. An interaction between the tempo and modality of the cues ($F_{1,18}=20.805, p<.001$) highlighted differences between the Auditory-Visual and

Visual-Only conditions for the Fast condition (*p<.001*), with addition of the auditory cue resulting in the participants' step intervals becoming closer to those generated by the avatar (Fig 2a). A similar result is seen for the Slow condition, but the change is not significant (*p=.084*; Fig 2b).

**Fig 2. Mean Inter-step intervals (ISI) of participant steps (filled bar) compared to the target Avatar ISI (lighter shaded bars) for all conditions.** Error bars show standard error of the mean (SEM).

Despite the difference between modalities, participants consistently had slightly longer ISI (slower steps) than the avatar in the Fast condition for both Auditory-Visual (*t(11)=9.635, p<.0001*) and Visual-Only conditions (*t(7)=7.989, p<.0001*). Similarly, participants continued to have slightly longer step intervals for the Slow tempo, but this was only significant for the Auditory-Visual condition (*t(11)=12.136, p<.0001*).

We found no significant differences in step interval variability across either the tempo or modality conditions.

## Synchronisation of steps

To analyse whether participants were able to synchronise their footsteps with the avatar, asynchronies were calculated between corresponding step onsets (see Fig 1 and Methods, Data Analysis section for how these were extracted). The asynchronies were then averaged across the trial up to the point of the perturbation, for each trial. Effects of each condition on the mean and standard deviation of asynchrony were analysed using a mixed measures ANOVA as before. The asynchrony was deemed negative when the participant led the avatar during a heel strike, and positive when the participant lagged behind as is commonly reported in the literature.

Mean asynchrony (Fig 3) was affected by both the avatar's stepping tempo ($F_{1,18}=8.533, p=0.009$) and whether the cues included both auditory and visual or just visual information ($F_{1,18}=11.510, p=0.003$). Moreover, an interaction between Tempo and Modality ($F_{1,18}=17.445, p=0.001$; Fig 3a) highlighted that while asynchrony remained consistent for fast tempos regardless of modality ($t(15.682)=-0.066$, $p>0.05$), it switched from a negative (Visual-Only) to positive (Auditory-Visual) asynchrony for the Slow condition ($t(15.325)=6.219, p<.0001$). Participants completing the Auditory-Visual condition showed similar positive mean asynchronies for both Fast and Slow tempos ($t(20.341)=-0.887, p>0.05$; Fig 3a), while those synchronising with the Visual-Only cues were on average ahead of the cue for the Slow tempo and lagged behind the cue for the Fast tempo ($t(13.148)=4.249, p<0.0001$; Fig 3a).

**Fig 3. Asynchrony between participant step onsets and corresponding Avatar steps a).** Mean Asynchrony between participant step onsets and corresponding Avatar steps, for all conditions. Negative asynchronies indicate the participants are, on average, stepping ahead of the Avatar cue. Error bars show SEM. **b).** Mean standard deviation of asynchrony for all conditions. Error bars show SEM.

Variability of asynchronies did not significantly differ between Fast and Slow tempos. However, there was a significant effect of Modality ($F_{1,18}=4.641, p=0.045$), with participants' timing being more stable with the addition of an auditory component to the cue (Fig 3b).

## Correction in response to temporal perturbations

Relative asynchrony was calculated, by subtracting the mean asynchrony prior to the perturbation from all resulting asynchronies in a trial. This measure eliminated the effects of individual participant tendencies to lead or lag behind the avatar when averaging across participants, with the relative asynchrony being approximately zero immediately prior to the perturbation event for each trial.

Participant asynchrony data was subsequently aligned so that the asynchrony corresponding to the perturbed step lay at step T. Plots for average asynchronies, from step T-4 to step T+6, across participants for each trial type are shown in Figs 4 and 5. For the Visual-Only conditions asynchronies tended to exhibit slow drift in the Fast Visual-Only condition, for both shortening and lengthening of steps, highlighting the lack of synchronisation between the participant and the avatar (Fig 4a,b). As the asynchrony continues to increase post-perturbation, it is assumed that there is no overall correction response to the perturbation for these conditions. For the Slow tempo, participants were able to maintain synchrony, shown by the asynchronies remaining close to zero. In particular, any drift seen in the Fast conditions does not continue to exacerbate post-perturbation. The effect of the perturbation can be seen in Fig 4c at time T, with a sudden increase in negative asynchrony, with the participant correcting back to stability by around T+3. High inter-subject variability for the lengthened step (Fig 4d) results in a less pronounced increase in positive asynchrony at time T and less clear correction strategy.

**Fig 4**. **Mean relative asynchronies before and after the perturbation (vertical grey bar) at time T for Visual-Only cues**. Dotted horizontal line shows the zero relative asynchrony measure, to which participants were expected to correct towards following the perturbation. Separate plots are shown for Fast, shortened **(a)** and lengthened **(b)** intervals, and Slow, shortened **(c)** and lengthened **(d)** intervals. Error bars show SEM.

**Fig 5. Mean relative asynchronies before and after the perturbation (vertical grey bar) at time T for Auditory-Visual cues**. Dotted horizontal line shows the zero relative asynchrony measure, to which participants were expected to correct towards following the perturbation. Separate plots are shown for Fast, shortened **(a)** and lengthened **(b)** intervals, and Slow, shortened **(c)** and lengthened **(d)** intervals. Error bars show SEM.

In contrast, for the Auditory-Visual condition, asynchronies were much more stable post-perturbation and inter-subject variability remained relatively low. There was a clear response following the perturbation in both fast and slow conditions. Participants corrected within 3 steps, with correction normally initiated after the second step post-perturbation for the Fast condition.

To quantify the temporal corrections to the perturbation for each condition, a linear phase correction model was fitted to the participants' asynchronies following the perturbation, using the method developed by Jacoby and colleagues[23]. Using the fitted model, the correction gain, $\alpha$, was calculated (Fig 6) representing the average proportion of the last asynchrony corrected for on the next step ($0<\alpha<2$; see Methods). These were then analysed using a mixed measures ANOVA to investigate the effects of each condition on the participants' ability to correct their timing in response to the perturbed step cycle.

**Fig 6. Mean correction gain as calculated by a linear phase correction model fitted to the asynchrony data of each participant**. Higher values indicate participants are correcting their movements to a bigger proportion of the preceding step asynchrony between themselves and the Avatar. Correction gains are shown for Fast versus Slow and Visual-Only (VO) versus Auditory-Visual (AV) conditions. Error bars show SEM.

Substantiating the visual results from the plots of relative asynchrony (Figs 4 and 5), participants showed lower levels of correction to the Visual-Only condition, compared to the Auditory-Visual condition ($F_{1,18}=9.323, p=0.007$; Fig 6). Additionally, participants showed much lower correction when the Avatar was stepping Fast, compared to Slow step intervals, indicating that they struggled to keep up and synchronise with the shorter stepping cycles ($F_{1,18}=19.885, p<0.001$). There were no interactions between the conditions and the direction of the perturbation had no significant effect on the correction gain indicating that participants corrected similarly when speeding up their steps to recover synchrony as they did when slowing down.

## Discussion

Here, we have investigated whether healthy participants are able to accurately coordinate their movements with a virtual avatar using a sensorimotor synchronisation paradigm. Participant's step tempo and asynchrony with respect to the avatar were measured while participants attempted to step in time with the avatar at two different step time intervals, 400ms (Fast) and 800ms (Slow). Correction gain was also calculated in response to temporal perturbations to one of the avatar's step cycles, both speeded and slowed by 15% at both tempos.

We found that the multimodal cues (visual with auditory foot-steps) reduced error in tempo, variability in asynchrony and increased participants' levels of temporal correction with the avatar. Participants were further able to synchronise more accurately and showed greater correction, when the avatar stepped at a slow pace. However, participant step onsets appeared to consistently lag behind the avatar's by approximately 150ms across both tempos and modal conditions.

A key finding from the experiments is the difference in participant performance between Visual-Only and Auditory-Visual conditions. The results from the stepping trials show that the addition of auditory cues significantly increased the accuracy of synchronisation and greater correction compared to visual-only cues. Moreover, in the visual-only condition, participants struggled to maintain synchrony and for the fast condition, even match the avatar's tempo. Multimodal cues have been found to increase accuracy in other movement timing contexts ranging from simple cues involving finger tapping[20] and larger full body movement such as gait[21]. Bimodal cues, featuring visual as well as auditory components, have been found to improve perception and synchronization to auditory rhythms[24] and reduce variability in asynchrony[25] over unimodal stimuli. In this study, we combined the complex spatio-temporal movements of the avatar with a simple relevant foot-strike sound effect equivalent to stepping on a gritted surface. This simple additional modality has provided the necessary sensory-information for participants to substantially improve their performance. The fact the cue is contextual to the action could also have contributed to the improved performance, with evidence that footstep sounds alone can be used to infer the kinematic and kinetic properties of walking[26]. While it could be argued that introducing the auditory cues could have resulted in participants ignoring the avatar's movements and exclusively using the discrete auditory events to synchronise their steps, there is overwhelming evidence in the literature that multisensory cues are integrated in a statistically optimal fashion described by maximum likelihood estimation[27], if the cues are relevant to each other[28],[29]. Hence, it is considered here that the improvement in timing performance is due to the visual avatar cues being integrated with the relevant foot step auditory cues.

Previously, sensorimotor synchronisation research with visual stimuli has focussed on simple cues such as flashing lights or bouncing spheres, which, while having significant effects on certain aspects of gait such as step timing and stride length[30], have limited scope for altering motion, given the amount of information they can convey. Existing research suggests spatio-temporal visual stimuli can result in improved synchronisation and coordination over temporal only stimuli as long as the spatial information is compatible[16],[17],[18]. In this study, participants failed to fully entrain to the spatio-temporal visual-only cues generated by the Avatar. This is highlighted further when comparing asynchrony results from this condition against existing literature assessing stepping on the spot synchronisation with metronome cues [21,31]. Mean asynchrony and SD of asynchrony were both significantly larger in our study: 120ms vs. -64ms[21] and -61ms[31] for mean asynchrony, and 100ms vs. 26ms[21] and 17ms[31] for SD of asynchrony. This could be due to the immersive nature of virtual reality headsets; participants lose out on visual feedback on their true surroundings and importantly their own limbs. This removal of sensory information could have made it difficult to align their own movements with the Avatar. This could be examined through the use of augmented reality (AR) avatars rather than virtual reality. AR would allow the participants to view their true environment and their own body while still getting a stereo view of the avatar.

Participants showed high deviation in terms of target tempo when attempting to match to fast cues, which resulted in the lack of synchrony. This subsequently increased variability when, based on previously observed Weber's law properties, lower variance should be observed for faster tempos[32]. Other studies have shown that participants are able to step in time with auditory and simple visual cues at an ISI of 500ms or above[21,31], closer to normal self-selected walking speeds, which indicates that fast tempos, such as 400ms, while reasonable for simple finger movements[33], is too fast for complex full-body movements without the addition of a corresponding auditory stimulus. In terms of the potential application of this study, the speed of the movement presented to participants should hence be considered in the context of training movements.

Participant's step onsets tended to lag behind the avatar by approximately 150ms. This asynchrony is opposite to that found in the majority of the tapping literature[33] but has been observed in some studies which involve movement over larger amplitudes[34]. The decision to use the step onset (such that we could observe the synchronisation in the start of the movement rather than the end) could also be a contributing factor to this observation; most finger tapping studies look at the moment of the tap – i.e. the end point of the movement when it hits a surface. Other studies into gait entrainment have shown that participants tend to synchronise with the peak velocity rather than a particular position within the gait cycle[35]. The velocity profile of the cue can affect coordination, with slowing to endpoints and turning points in trajectory increasing the stability of coordination[36]. This highlights the potential for movement alignment to vary across the trajectory and future studies will need to look at the temporal-spatial matching of the participants' trajectories to that of the avatar, rather than discrete time points.

Responses to perturbations in the avatar's step cycle were clearly related to participants' synchronisation performance. Participants were unable to correct their timing when attempting to synchronise with fast stepping on the spot cues without the aid of footstep sounds and, in general, participants corrected better to slow rhythms. Correction gain also increased significantly in the presence of auditory cues, with participants on average correcting within 2 to 3 steps after the perturbation. This is in agreement with existing literature, which states that the bimodal condition lowers variability in asynchrony and provides an advantage over visual only cues when correcting to phase perturbations[25]. This indicates that if participants are able to keep in sync they have

'confidence' in the cue and hence make larger and quicker corrections, corroborating results from Wright, Spurgeon & Elliott[21]. In their study, Wright et al. found that mean correction to phase perturbed metronome cues, calculated as the percentage correction compared to the initial perturbation, ranged from 7.5% in the Visual-Only case to 17% Audio-Visual, at an ISI of 500ms, when averaged across the 5 steps post-perturbation. Additionally, Bank and Roerdink's study[13], assessing responses when treadmill walking at self-selected speeds, found an average correction of 18% in the 5 steps following a perturbation and Chen, Wing and Pratt[31] found a mean correction at step T+1 of 34.4% at an ISI of 500ms, which would be expected to reduce in subsequent steps. In our study, at 400ms, we found an equivalent correction response in the audio-visual condition. At 800ms, however, we found a significantly better correction response, with correction gain, α, ranging from 0.265 to 0.405, equivalent to 26.5% and 40.5% if the correction is linear, when calculated using the method applied by Wright et al. [21], in the Visual-Only and Audio-Visual conditions respectively. This suggests that, barring the advantage afforded by the longer period, multisensory cues in immersive Virtual Reality can be used as exercise cues for full body motion to influence a person's own movement.

## Conclusion

In conclusion, in this study, we have demonstrated that humanoid cues within virtual reality can be used as exercise cues for entrainment, as participants are able to match and correct the timing of their movements to an avatar. However, our study has shown that it is important to consider the use of multimodal cues along with relatively slow movement tempos to achieve the best level of entrainment.

These results open up the use of virtual avatars for rehabilitation, allowing physiotherapists to present realistic exercises to patients with an 'ideal' leader to follow. These avatars provide patients with natural cues, taking advantage of their ability to mirror the motion of other humans. Additionally, providing a humanoid cue which does not vary between repetitions of an exercise allows for participant motion to be analysed against any target motion, aiding in the identification and quantification of deficiencies in a participant's limb function.

# Materials and methods

## Participants

Twenty, healthy participants (male, right-handed, aged 23-39) took part, and completed, the experiments. None reported any lower limb function disabilities. Participants were unaware of the purpose of the experiment and gave written informed consent prior to taking part. One participant reported having taken musical training (drums) but the remainder did not have any musical or dance related experience. Eight participants completed the first experiment ("Visual-Only condition"); twelve completed the second experiment ("Auditory-Visual" condition). As we were primarily interested in the effects of avatar/visual cues, Audio-only trials were not conducted. Instead, results from this study's visual-only and audio-visual trials were compared with equivalent audio-only trials and trials involving simple cues.

The study was approved by the University of Warwick Biomedical Sciences Research Ethics Committee (BSREC), reference REGO-2015-1582, and was conducted in accordance with the ethical standards of the 1964 Declaration of Helsinki.

## Design

Each experiment consisted of a single session per participant which lasted approximately 40 minutes. In both experiments participants wore an Oculus Rift DK2 (Oculus VR, California, USA) virtual reality headset to view an avatar in a virtual environment (Fig 7), developed in Unity3D (Unity Technologies SF, California, USA).

**Fig 7. View of the avatar seen by participants when completing the stepping task.**

During the session, 40 reflective markers were attached to the participants, placed using the "plug-in-gait"[22] model for capturing full body motion. Participants stepped on the spot inside a 3m by 4m marked space and their movement was tracked and recorded by a 3D motion capture system (Vicon Vantage, 12 cameras) at a sampling rate of 100Hz.

Participants were instructed to step on the spot in time with the avatar (see Stimuli below), which was viewed on the headset. The orientation was designed such that the participant stood behind the avatar, with the whole body in view, this was to ensure the avatar's feet were not hidden as occlusion of end-points have been found to increase variability in coordination[37]. Participants completed either Experiment 1 (Visual-Only) or Experiment 2 (Auditory-Visual) which involved 4 blocks of 5 trials (20 trials in total) run in succession. Short breaks were offered between blocks, every 10-15 minutes, additional breaks could also be requested by participants. Each trial consisted of the participant synchronising a total of 30 steps with the Avatar.

One interval from step 10-16 was chosen to be perturbed. The intervals pre-perturbation were used for the ISI and asynchrony analyses, with the first 3 steps were excluded. The intervals post-perturbation were used for the correction gain analysis. More information about the trials and analysis is given later in this section.

The experiment was a 2 (Modality: Visual-Only, Auditory-Visual) x 2 (Tempo: Slow, Fast) x 2 (Perturbation Direction: Positive, Negative) design. Tempos represented the time between step onsets, denoted as the inter-step intervals. For the slow condition, this was approximately 800ms, while the fast condition was approximately 400ms. Step intervals generated by the avatar were not exactly at the target value as the avatar movement data was generated by real human motion to a metronome set to the target intervals. Therefore, some variation due to this and frame rate of the headset occurred.

Participants who completed Experiment 1 followed the avatar without any auditory cues, i.e. Visual-Only. This was performed in a quiet room with a relative soft surfaced floor, so auditory feedback from their own footsteps was minimal. Participants who completed Experiment 2 completed the experiment with the same visual cues but with footstep sounds (shoe onto gritty/dirt surface) accurately overlaid to align with the heel strikes of the avatar.

For all trials one step cycle of the avatar was randomly chosen (between steps 10 and 16) to be temporally shortened or lengthened by 15%, creating a perturbation. On the whole participants were unaware of this perturbation occurring.

Participants were not trained with, or acclimatised to, the system before the session, but did receive one practice trial to familiarise with the task. The order of the within-subject conditions were randomised to cancel out any potential learning effects which may have occurred over the course of the session.

## Stimuli

The avatar cue was developed by recording marker trajectories on a volunteer (Male, 37 years) using the same Vicon motion capture system and full body market set used to collect participant data. The volunteer was instructed to step on the spot in time with an auditory metronome at either a fast (metronome interval: 400ms) or slow (metronome interval: 800ms) tempo. These tempos were chosen to ensure participants had to actively attempt to stay in synchrony with the avatar rather than stepping at a speed close to what is reported as a typical self-selected pace (545ms[38]).

The virtual environment was developed using the Unity3D game engine. The scene presented in the application was chosen to be minimal, with no background movement or objects, to avoid distractions while the trials were being completed as high density environments have previously been found to affect step kinematics in walking experiments[39].

Humanoid characters were obtained from Adobe Mixamo (Adobe Systems Incorporated, California, USA). A male character was selected to represent the male volunteer who provided the movement data to be mapped on to the avatar. The motion data was imported into Unity3D, and a single cycle of stepping on the spot, which matched closest to the original metronome cue, was mapped onto the avatar. This was done to minimise the temporal and spatial variability in the exercise cue, as variance in an exercise cue is known to adversely affect sensorimotor synchronisation, increasing the variance, and phase correction, reducing the correction gain[40].

When viewing through the headset, participants were positioned behind the avatar who was facing away from them. This was representative of tasks requiring a "follow the leader" approach, such as dance classes, minimising the incongruence of the stimuli presented. This view also allowed users to easily internalise the motion of the avatar and compare their own motion to it, as their orientation matched that of the avatar[41]. Orientation of the participant's view was obtained directly from the virtual reality headset used to present the scene to reduce the lag between user motion in the physical and virtual worlds.

The trajectories of the avatar limbs from the end of the chosen step cycle to the start were interpolated to allow the step cycle to play repeatedly without jittering. A single step cycle was chosen instead of an average of multiple step cycles to keep the cue motion as natural as possible.

To create the Perturbation, one step cycle, between the 10$^{th}$ and 16$^{th}$ step was chosen at random, and either accelerated (resulting in a shortened step interval, or negative perturbation) or decelerated (resulting in a lengthened step interval, or positive perturbation) by 15%. Accelerated or decelerated intervals were achieved by changing the playback speed during the step interval period. The step chosen randomly to be perturbed was achieved via a script within Unity3D.

Natural footstep sounds, representing a shoe on a gritty surface, were obtained from the Unity Asset Store and synchronised with the avatar heel strikes for the second Auditory-Visual condition. The phenomenon of spatial ventriloquism[42], where temporally synchronous sound is perceived to occur at the location of a visual stimulus, meant that the origin of the sound did not need to be placed at the base of the avatar in order for participants to perceive it coming from the feet.

## Procedure

A hardware trigger was developed to synchronise trial recordings from Vicon with the recording of the exercise cue within Unity3D. Within the application, once a trial was started, which triggered recoding of the avatar motion within Unity3D, a rising edge was sent to the computer's serial port. This was connected to the Vicon system which was configured to start recording when a rising edge was received, thereby synchronising the start times of the recordings from the two systems.

The scene, including the trial animations, logic for selecting trials and perturbing steps, was driven by the Unity3D application running on a Windows 7 desktop PC. This was presented to the participants through the Oculus Rift DK2 headset at 75Hz, with a resolution of 960x1080 pixels per eye.

Time-stamped trajectories for all limbs were captured by both the Vicon and Unity3D systems, for the participant and avatar, respectfully. As each system captured at a different rate, the Vicon system capturing at 100Hz versus the Unity3D/Oculus system recording at approximately 75Hz, the recordings were synchronised by timestamp, rather than frame, offline in MATLAB.

Participants self-initiated each trial by pressing a key positioned on the floor with their foot, after which the avatar started playing the exercise cue. Participants were instructed to step in time with the avatar, matching left and right leg movement. Participants were not informed of the presence of a perturbation in the step cycle.

## Data Analysis

The stepping data was analysed by extracting the left and right step onset times of both the participant and the avatar for every trial. These were extracted in MATLAB by identifying the time point when the Y axis coordinate of the captured heel marker rose above a predefined threshold height above the floor surface. Participant heel strike times were temporally aligned to the avatar onsets, allowing for calculation of timing errors, or asynchronies, between each heel strike and the corresponding cue. The duration between heel strikes was used to measure the inter-step interval (ISI) for comparison of tempo.

In the case where the participant ISI was consistently different from the avatar ISI, phase drift occurred, where participant motion moved from being in-sync to anti-phase to in-sync in later steps causing a wrapping effect in the asynchrony (as, initially, the closest steps were chosen as being corresponding). When this was observed, corresponding steps were assigned to ensure continuous asynchrony between steps. In certain cases, this meant asynchronies appeared to grow larger than the cue's inter-step interval but this was required to correctly calculate mean asynchrony across steps (Fig 8).

**Fig 8. Asynchronies before and after phase wrapping removed a).** Asynchronies observed for a fast tempo trial when matching participant onsets to the nearest Avatar step onsets. Wrapping of the asynchrony can be seen between the intervals of +/-0.4 seconds which does not correspond to an attempt to regain synchrony. **b).** Asynchronies for the same trial but with phase wrapping removed and steps assigned to ensure asynchronies are continuous.

After the unwrapping of asynchronies, a total of 7 fast tempo trials out of 76 for the Visual-only condition and 2 out of 120 or the Audio-visual condition were excluded from further analysis due to multiple steps being missed. Similarly, for slow trials, 2 out of 62 from the Visual-Only and 1 out of 120 from the Auditory-Visual condition were removed.

Mean asynchrony and ISI were calculated pre-perturbation, from step 3 to the perturbed step, for each participant within each condition. These were subsequently averaged across participants. The first 3 steps were eliminated from the analysis to remove the steps participants used to initially synchronise with the avatar after the abrupt start as it has been found that there is significant variability between subjects in time to synchronisation, and working memory plays an important role in in synchronisation during the initiation phase[43]. This resulted in between 7-13 steps per trial being used for the inter-step interval and asynchrony analyses pre-perturbation.

Means from the recording of the avatar during the trial were used in the statistical analyses instead of the original metronome tempo as the avatar was subject to the natural variance of the original volunteer. Dropped frames and decreases in frame-per-second (FPS) rates when the application was running also had an effect on the final tempo of the avatar observed by the participants, which varied between trials.

A repeated measures ANOVA was used to test for significant differences in mean ISI and asynchrony between conditions, auditory-visual versus visual cues and fast versus slow tempos, and paired t-tests were used to determine whether means and standard deviations of the asynchrony and inter-step interval were significantly different from the avatar cue.

Relative asynchronies were calculated by obtaining the mean asynchrony for all steps pre-perturbation per trial type per participant and subtracting these from the steps post-perturbation, to eliminate effects the of individual participant tendencies to lag or lead the avatar. The relative asynchronies, for each step post perturbation, were used to investigate the corrective response. This was achieved by fitting a linear phase correction model[23,44,45] to the asynchronies, based on the recurrence equation shown in Eq. 1:

$$A_{n+1} = (1-\alpha)A_n + T_n + M_{n+1} - M_n - C_n \quad (1)$$

where $\alpha$ is the correction gain, $A_n$ is the current event asynchrony, $T_n$ is the time interval generated by an assumed internal timekeeper, $M_n$ is the current motor implementation delay and $C_n$ is the corresponding step interval time of the avatar. A correction gain of 1 is therefore full correction of the previous asynchrony, while a correction gain of 0 indicates no correction.

The bounded General Least Square approach developed by Jacoby et al[23] was used to estimate the parameters of the model fitted to each trial. In particular, we extracted the correction gain, $\alpha$, to determine the level of correction participants exhibited under each of the conditions. Correction gain values obtained per participant per trial type were analysed using a Repeated Measures ANOVA to find significant changes between conditions.

## Data Availability

The datasets generated during and/or analysed during the current study are available at [REPO LOCATION].

## Acknowledgements

All contributors to the work have been acknowledged as co-authors .

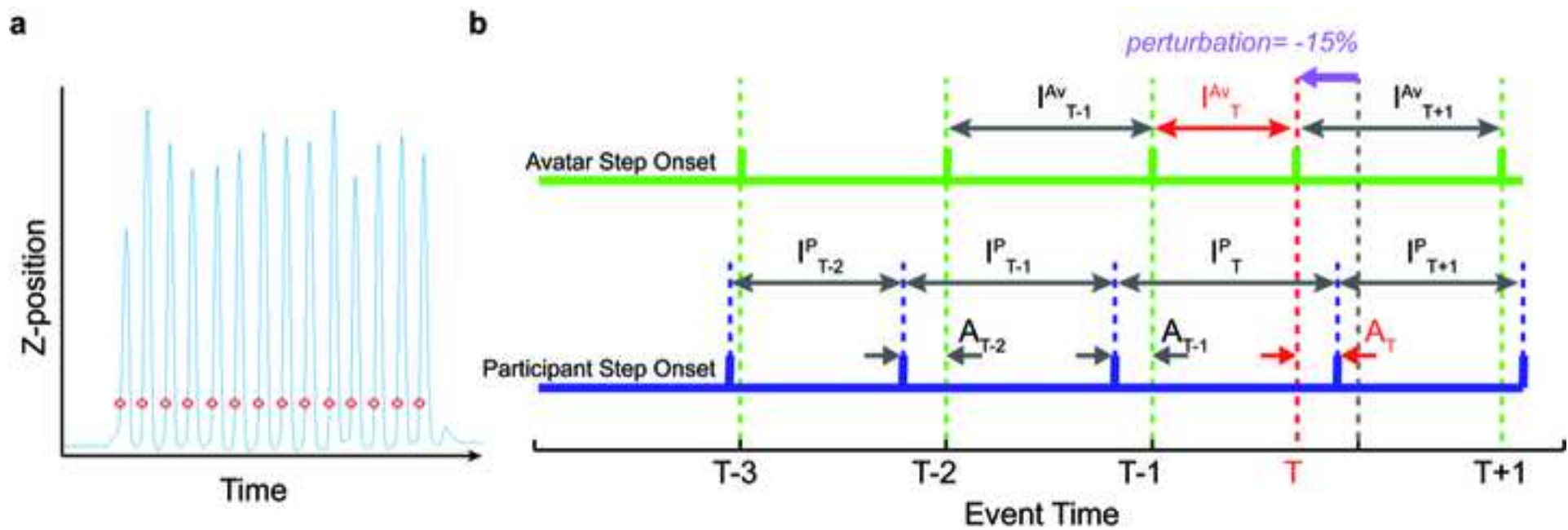

Figure 2

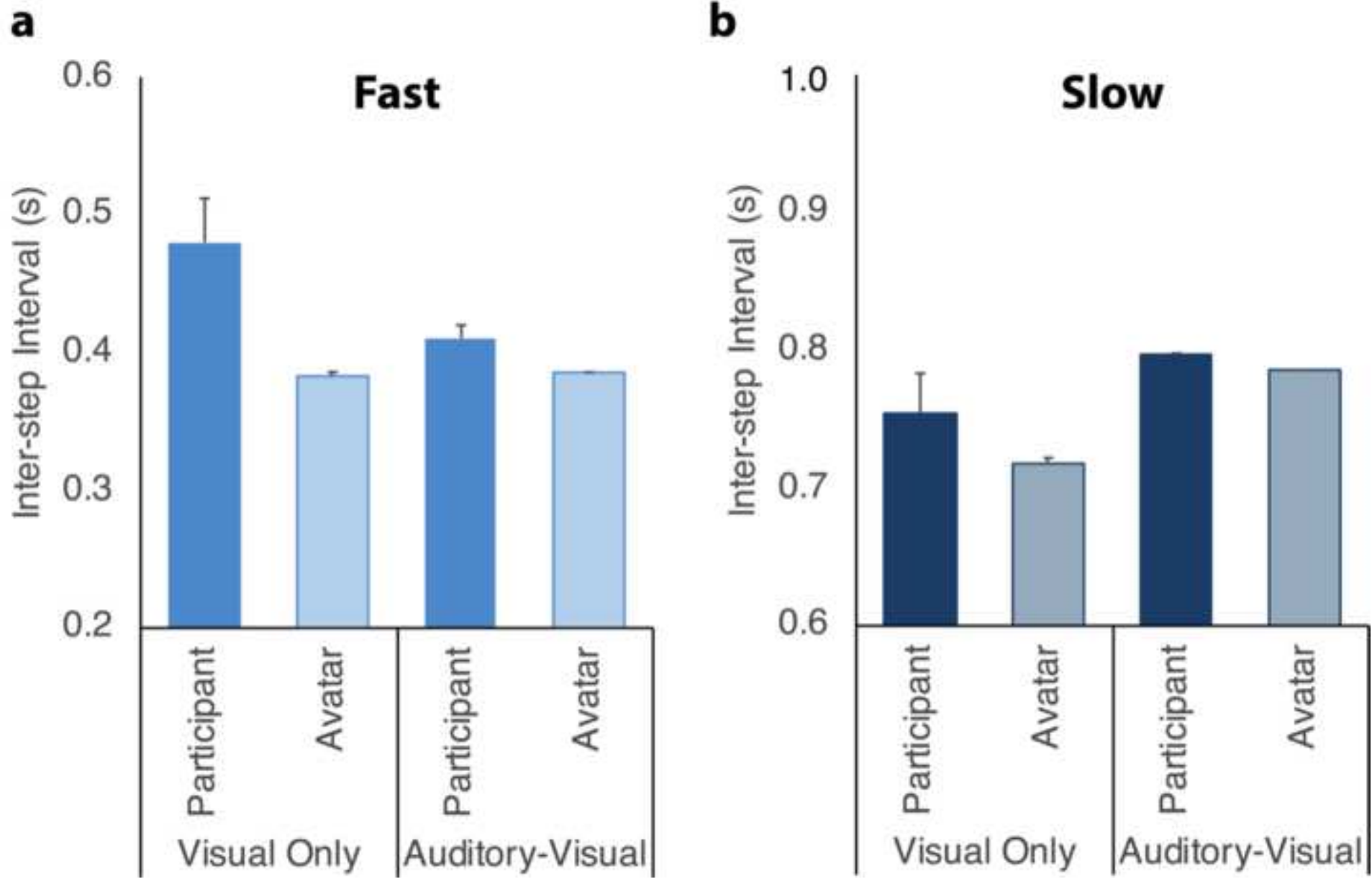

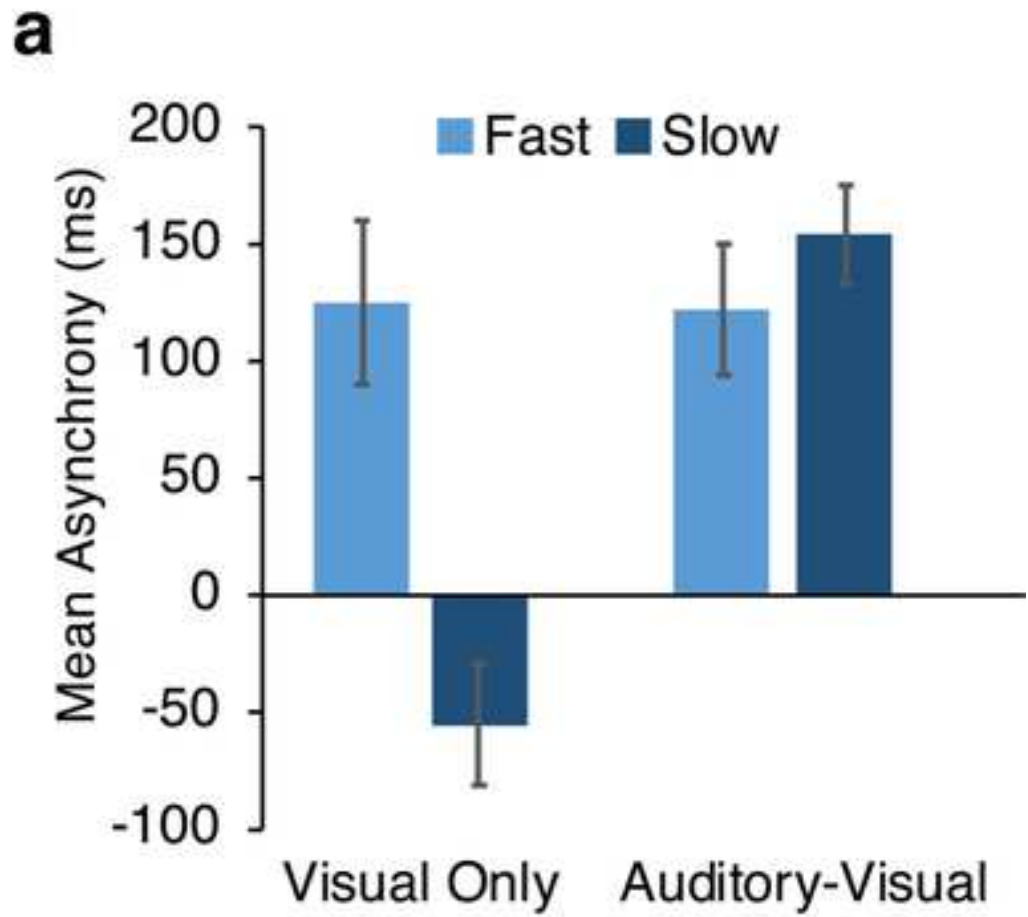
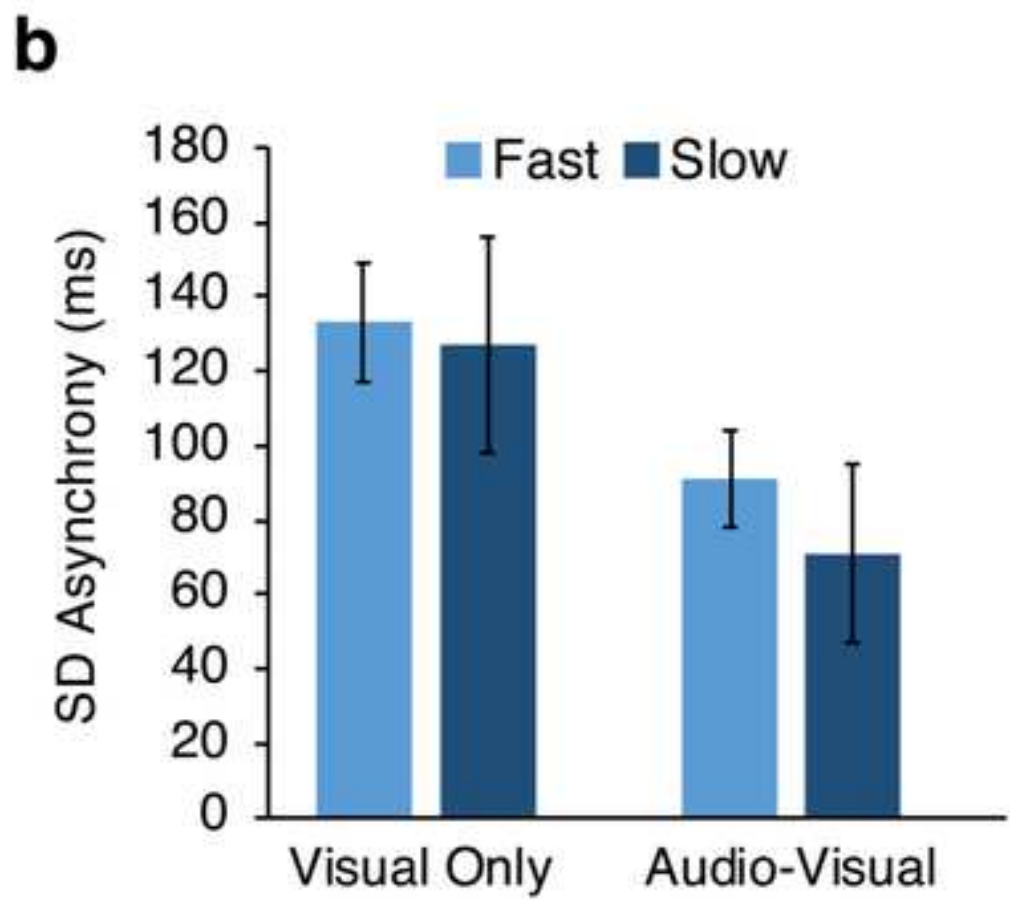

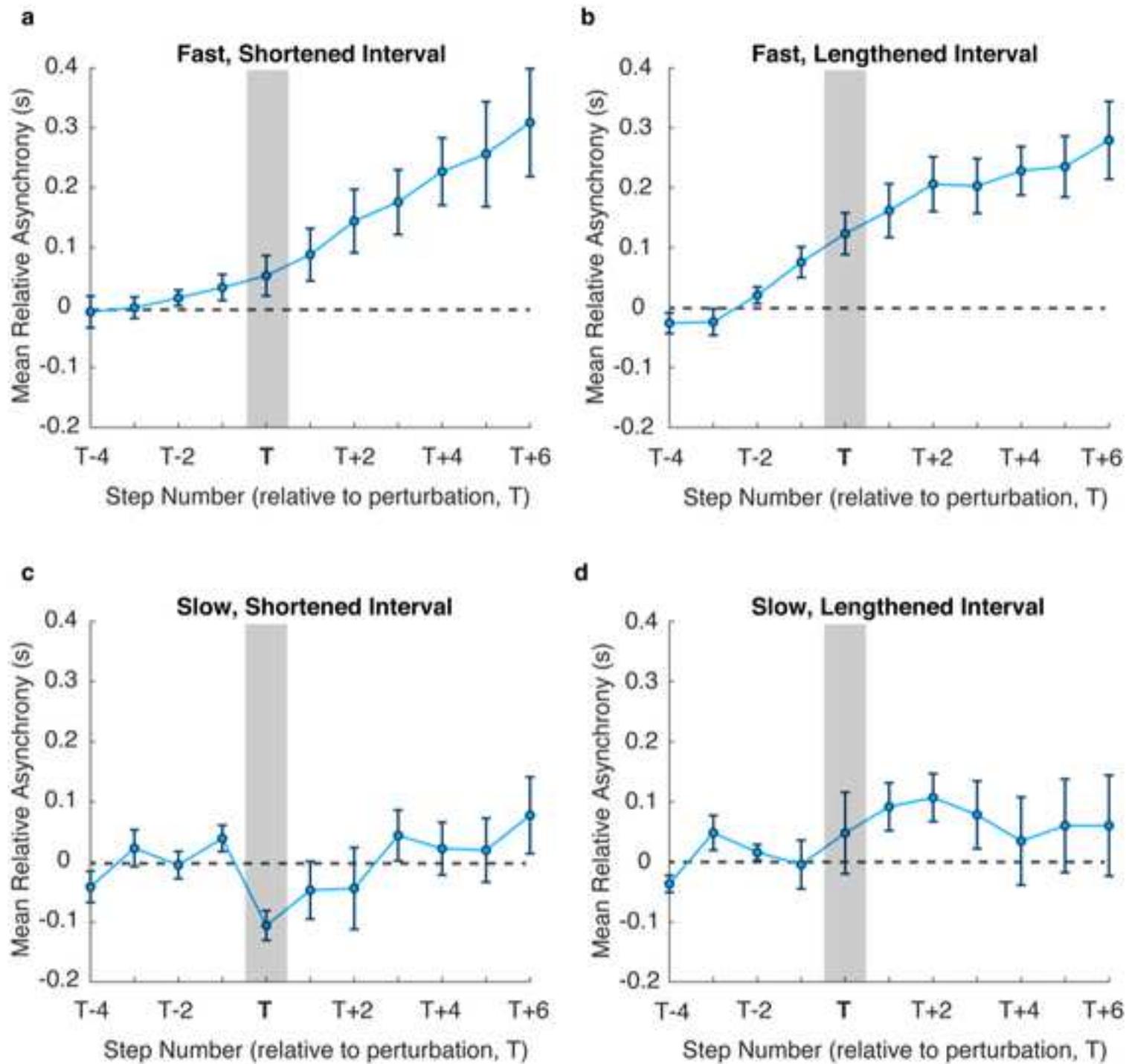

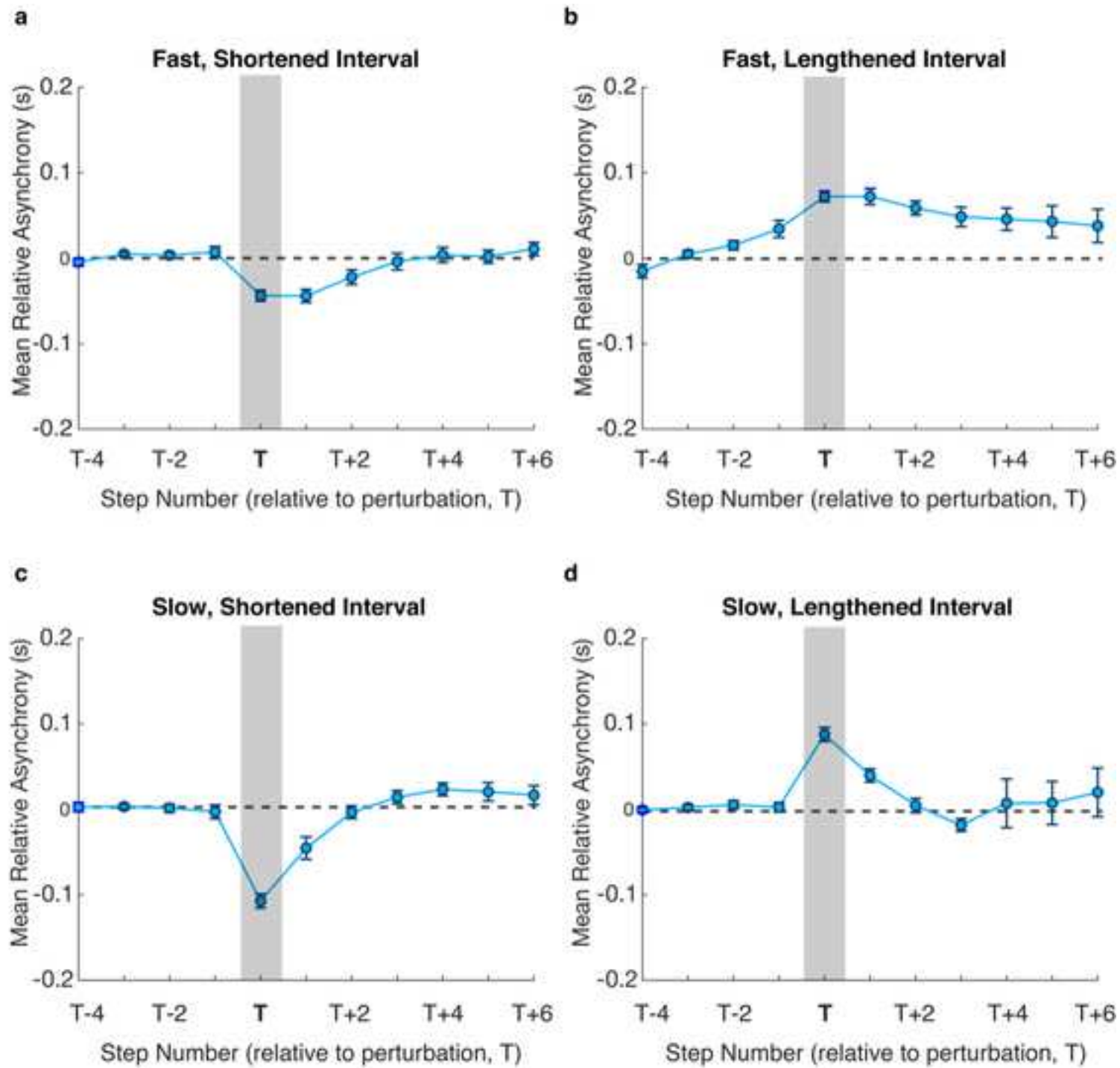

Figure 6　　　　　　　　　　　　　　　　　　　　　　　　　　　　　Click here to access/download;Figure;Fig6.tiff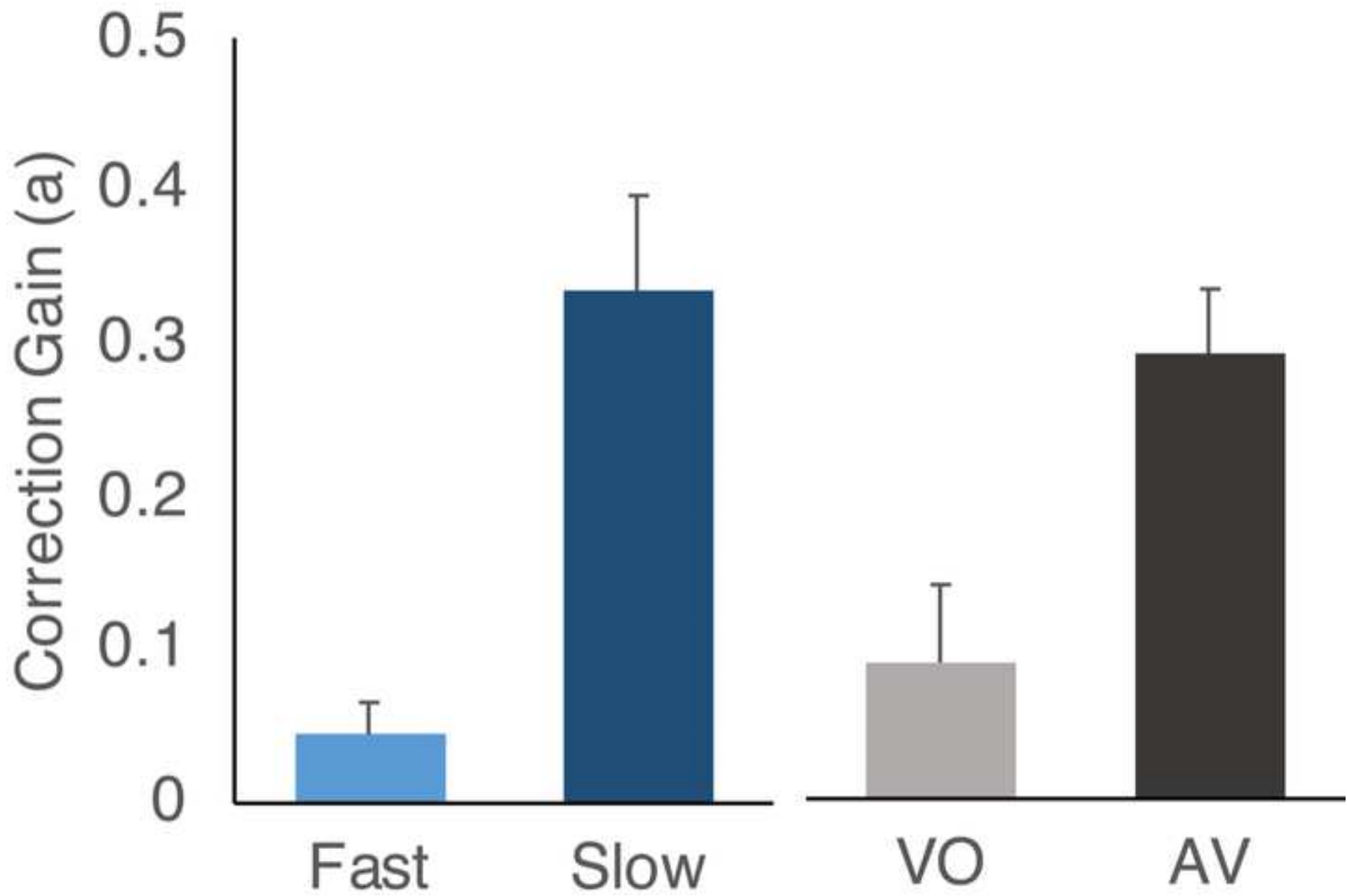

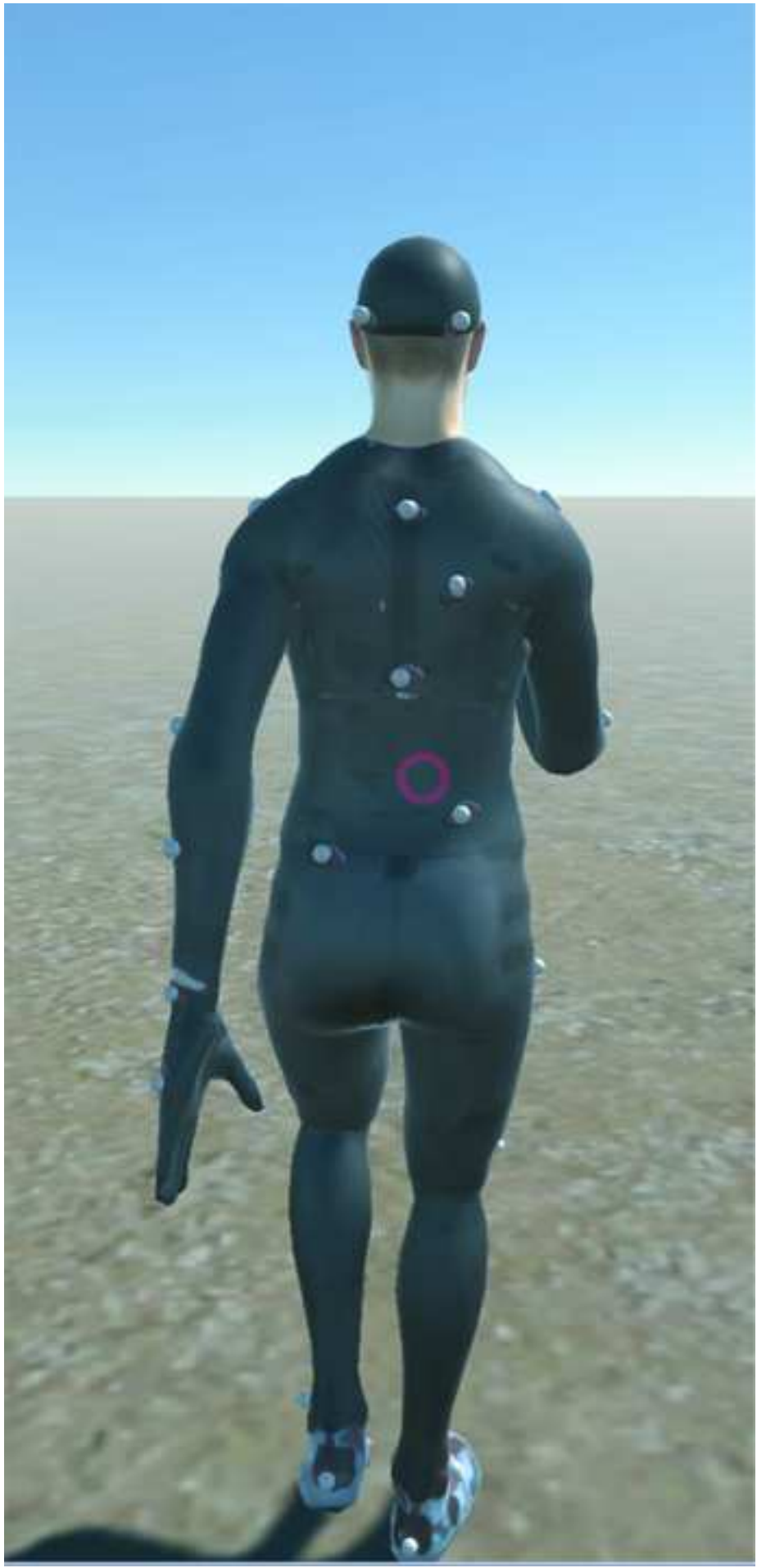

Figure 8

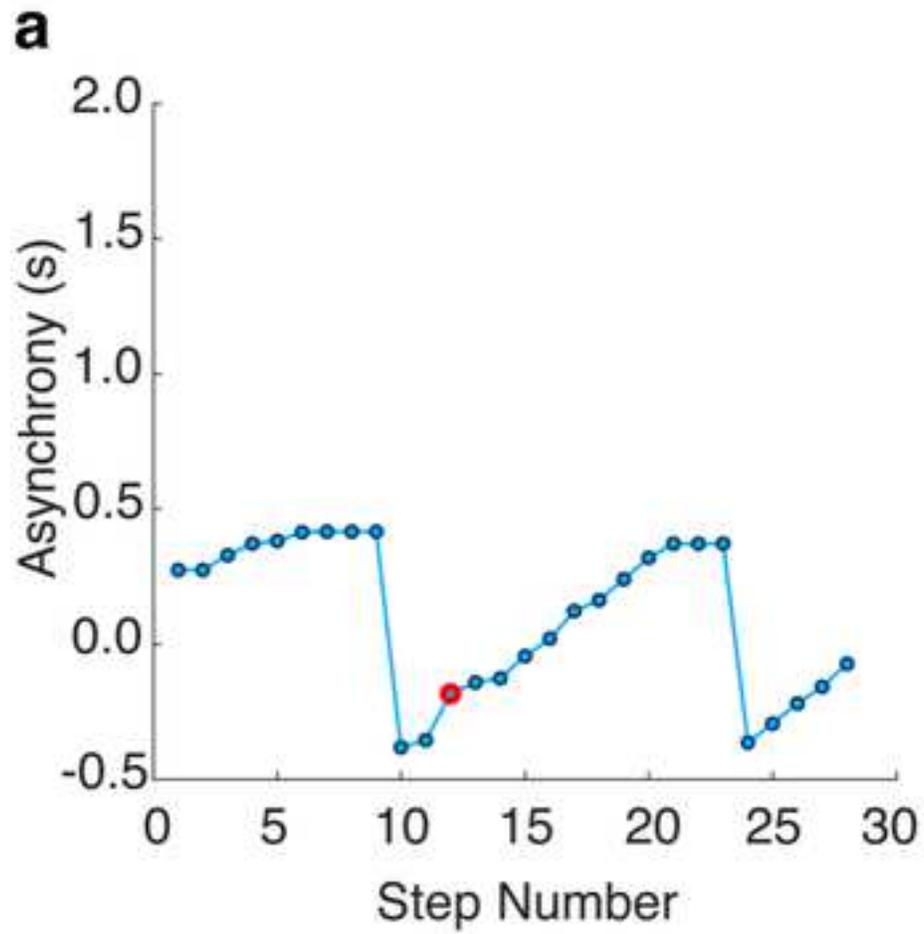

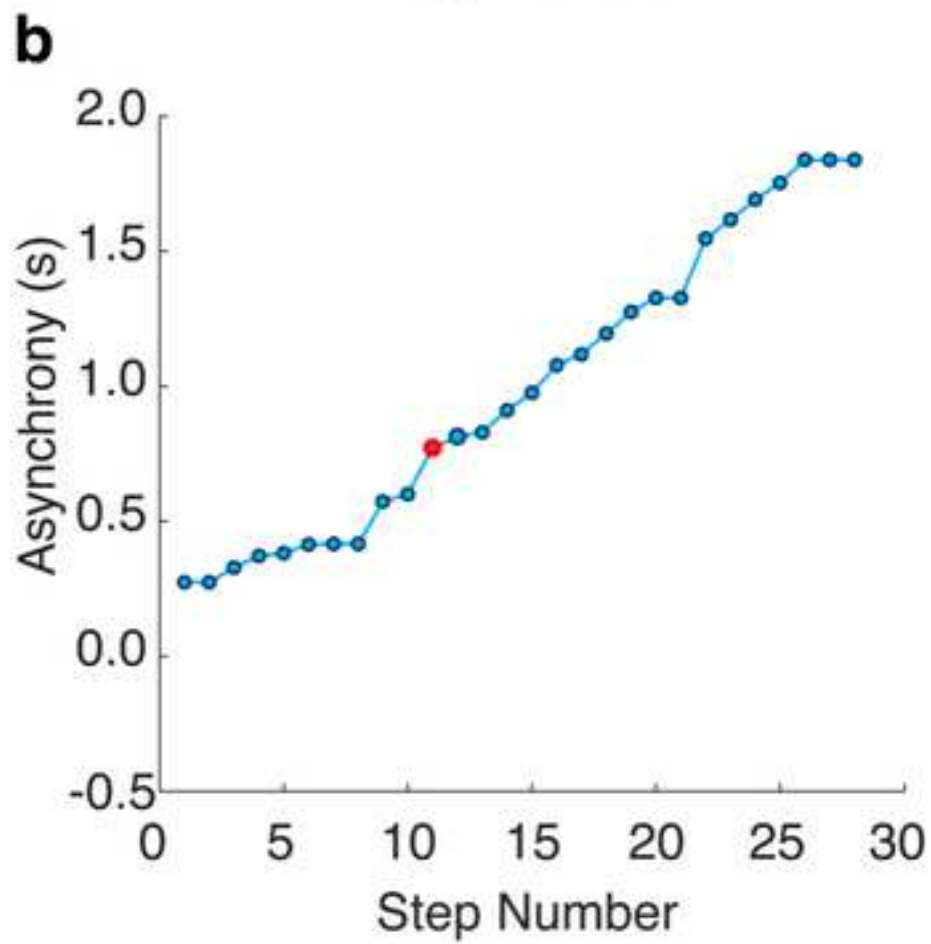